\documentclass[aps,prl,twocolumn,superscriptaddress]{revtex4-1}
\usepackage[USenglish]{babel} 
\usepackage[T1]{fontenc}
\usepackage[ansinew]{inputenc}

\usepackage{lmodern} 

\usepackage{graphicx} 
\usepackage{amsmath}
\usepackage{amsthm}
\usepackage{amsfonts}
\usepackage{color}
\usepackage{subcaption}

\begin{document}

\title{Giant catalytic effect of altruists in Schelling's segregation model}

\author{Pablo Jensen}
\email{corresponding author, pablo.jensen@ens-lyon.fr}
\affiliation{Institut Rh\^{o}nalpin des Systemes Complexes, IXXI, F-69342 Lyon, France}
\affiliation{Universite de Lyon, Laboratoire de Physique ENS Lyon and CNRS, 46 Rue d'Italie, F-69342 Lyon, France}

\author{Thomas Matreux}
\affiliation{Institut Rh\^{o}nalpin des Systemes Complexes, IXXI, F-69342 Lyon, France}
\affiliation{Universite de Lyon, Laboratoire de Physique ENS Lyon and CNRS, 46 Rue d'Italie, F-69342 Lyon, France}
\affiliation{Physics Department, Ludwig-Maximilians-Universit\"at M\"unchen, 80799 M\"unchen, Germany}

\author{Jordan Cambe}
\affiliation{Institut Rh\^{o}nalpin des Systemes Complexes, IXXI, F-69342 Lyon, France}
\affiliation{Universite de Lyon, Laboratoire de Physique ENS Lyon and CNRS, 46 Rue d'Italie, F-69342 Lyon, France}

\author{Hernan Larralde}
\affiliation{Instituto de Ciencias F\'isicas, UNAM, Av. Universidad s/n, CP 62210 Cuernavaca Morelos, M\'exico}

\author{Eric Bertin}
\affiliation{LIPHY, Univ. Grenoble Alpes and CNRS, F-38000 Grenoble, France}

\begin{abstract}
We study the effect of introducing altruistic agents in a Schelling-like model of residential segregation. We find that even an infinitesimal proportion of altruists has dramatic catalytic effects on the collective utility of the system. Altruists provide pathways that move the system away from the suboptimal equilibrium it would reach if the system included only egoist agents, allowing it to reach the optimal steady state.
\end{abstract}

\maketitle
	
Simple social models may be useful to improve our intuitive, often implicit, conceptualizations of 
social processes \cite{rmp,watts,seuil}. For example, the segregation model proposed by Schelling \cite{schelling_dynamic_1971} helps understanding that the collective state reached by agents may well be different from what each of them seeks individually. Specifically, Schelling's model shows that even when all agents share a preference for a mixed city, the macroscopic stationary state may be segregated \cite{pnas}. In this Letter, we show that introducing a vanishingly small concentration of altruist agents gives rise to  a strongly non linear response, 

Our model combines two important themes for many disciplines, including physics and economics: 
The large effects of small perturbations and the influence of altruistic behavior on coordination problems. 
On the first point, microscopic causes leading to macroscopic effects are well-known in physics. 
Chaos theory has shown that some dynamical systems are prone to an exponential increase of small perturbations \cite{chaos}, a topic of recurring interest in other fields, such as modeling of ecological competition \cite{hebert} or pattern formation \cite{pattern}. More related to our work, there are several examples of large effects arising from small changes in population composition. It has been shown that a small variation in the proportion of uninformed individuals may lead to strong changes in the way collective consensus is achieved by animal groups manipulated by an opinionated minority \cite{couzin}. In the minority game \cite{minority}, introducing a small proportion of fixed agents - i.e. agents that always choose the same option - induces a global change in the population behavior, leading to an increase of the overall gain \cite{liaw1,liaw2}. In the voter model, a finite density of voters that never change opinion can prevent consensus to be reached \cite{redner}. 

On the second point, altruism is a major topic in evolutionary biology and economics \cite{fehr,boyd,kirman}. Many models have shown that pair interactions between selfish players lead to stationary states of low utility. They have introduced various types of altruistic behavior to investigate how it may lead to a better equilibrium: altruistic punishment \cite{fehr,boyd}, inequity aversion \cite{hetzer}, fraternal attitudes \cite{szabo}, agent mobility\cite{cong} \ldots Here, we use a simple definition of altruism (see below) and concentrate on the proportion of altruists needed to reach the social optimum. We show that, unexpectedly, an infinitesimal proportion of altruists can coordinate a large number of egoists and allow the whole system to reach the social optimum.

\paragraph{Description of the model.}

\begin{figure}%
\centering
\includegraphics[width=.65\columnwidth]{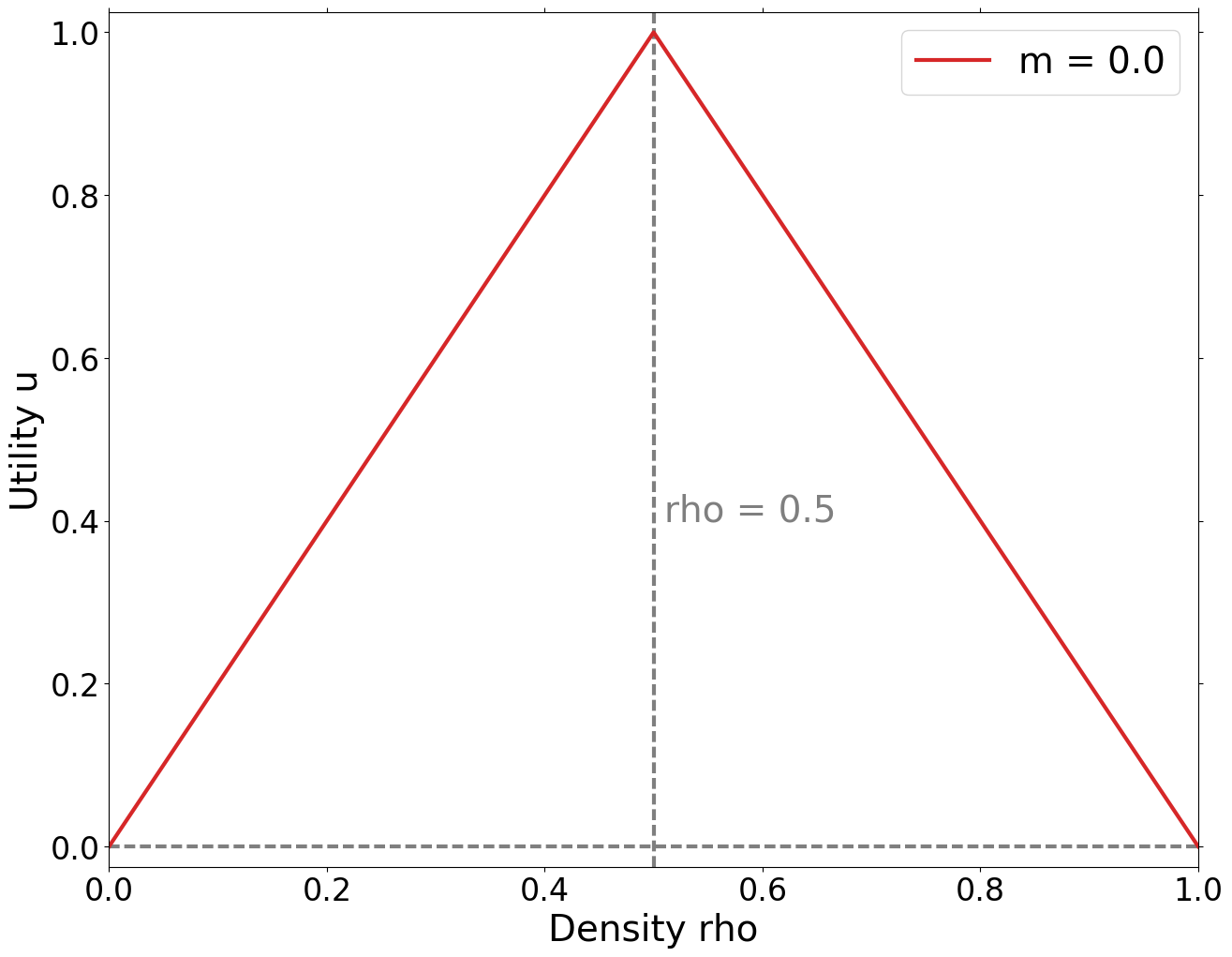}%
\caption{Agent utility function: $u(\rho) = 2 \rho$ for $\rho \leq 0.5$ and $u(\rho) = 2 (1 - \rho)$ for $\rho > 0.5$.}%
\label{fig:utility_function}%
\end{figure}

Our model represents the movement of a population of agents in a "city", which is divided into $Q \gg 1$ non overlapping blocks, also called neighborhoods. Each block is divided into $H$ sites and has the capacity to accommodate $H$ agents (one per site). Initially, a number of agents $N = Q H \rho_0$ are distributed randomly over the blocks, leading to an average block density $\rho_0$ ($\rho_0 = 0.4$ throughout the paper). All agents share the same utility function $u(\rho)$ that depends on the agents density $\rho$ in the neighborhood where they are located. We choose a triangular utility (see Fig.~\ref{fig:utility_function}): agents experience zero utility if they are alone ($\rho = 0$) or in full blocks ($\rho = 1$), and maximum utility $u=1$ in half-filled blocks ($\rho = 0.5$). The collective utility $U$ is defined as the sum of all agents' utilities,
$U = H \sum_{q=1}^Q \rho_q u(\rho_q)$ and the average utility $\tilde{u}$ per agent is $\tilde{u}= U / N$.

Building upon our past work on Schelling's segregation model \cite{pnas}, we now mix two types of agents: "egoists", who act to improve their own, individual, utility, and a fraction $p$ of "altruists", who act to improve the collective utility. Thus, egoists have as objective function the variation of their individual utility $\Delta u$, while altruists consider the variation of the overall utility $\Delta U$.  The dynamics is the following: at each time step, an agent and a free site in another block are selected at random. The agent accepts to move to this new site only if its objective function {\it strictly} increases (note that the moving agent is taken into account to compute the density of the new block). Otherwise, it stays in its present block. Then, another agent and another empty site are chosen at random, and the same process is repeated until a stationary state is reached, i.e., until there are no possible moves for any agent. 

\paragraph{Limiting cases: pure egoist or altruist populations.}
In \cite{pnas}, we have computed analytically the stationary states of a {\it homogeneous} population of egoist or altruist agents. Altruists always reach the optimal state, given by half filled (or empty) blocks and an average pure altruist utility $\tilde{u}_A \simeq 1$. In contrast, a pure egoist population {\it collectively} maximizes not $U$ but an effective free energy that we have called the {\it link} $L$. The link is given by the sum over all blocks $q$ of a potential $l_q$: $L = \sum_q l_q$, where $l_q = \sum_{n_q = 0}^{N_q} u(n_q/H)$, with $N_q= H  \rho_q $ is the total number of agents in block $q$. In the large $H$ limit, 
\begin{equation} \label{eq:def:link}
l(\rho_q) \approx H \int_0^{\rho_q} u(\rho) \, d\rho. 
\end{equation}
The link may be interpreted as the cumulative of the individual marginal utilities gained by agents, as they progressively enter the blocks from a reservoir of zero utility. Its key property is that, {\it for any move}, $\Delta L=\Delta u$. Since egoists move only when their individual $\Delta u$ is positive, the stationary state is given by maximizing $L$ over all possible densities $\{\rho_q\}$ of the blocks, from which no further $\Delta u > 0$ can be found. Analytical calculations \cite{pnas} show that this stationary state corresponds to crowded neighborhoods, far above the state of maximum average utility given by $\rho_q = 1/2$. 
For the case studied in this paper, the stationary density is given by $\rho_E = 1/\sqrt{2}$, leading to a pure egoist utility $ \tilde{u}_E = 2 (1 - \rho_E) \simeq  0.586 \ll 1 $. Numerical simulations have confirmed these results, 
though the existence of many metastable states around $\rho_E \simeq 0.7$ leads to fluctuations in the simulated final densities.

\begin{figure}%
\centering
\includegraphics[width=.85\columnwidth]{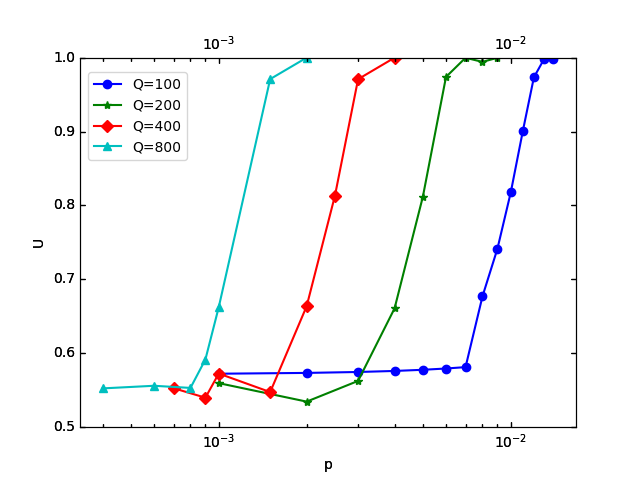}
\includegraphics[width=.85\columnwidth]{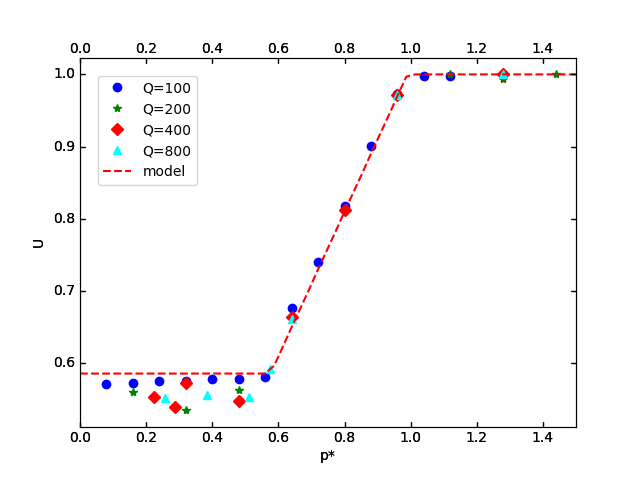}
\caption{Evolution of the average utility as a function of (a) the altruists' fraction $p$ (note the log scale on the x-axis) and (b) the rescaled fraction $ p^* = 2 p Q \rho_0$. We take $H=200$ and vary $Q$ as shown. The fluctuations for low $ p^*$ values (before the transition) arise from metastable states in the pure egoist regime.}
\label{fig:palpha}
\end{figure}

\paragraph{Mixing populations: qualitative picture.}

\begin{figure}[!htb]%
\begin{subfigure}[b]{.4\columnwidth}
\includegraphics[width=1.2\textwidth]{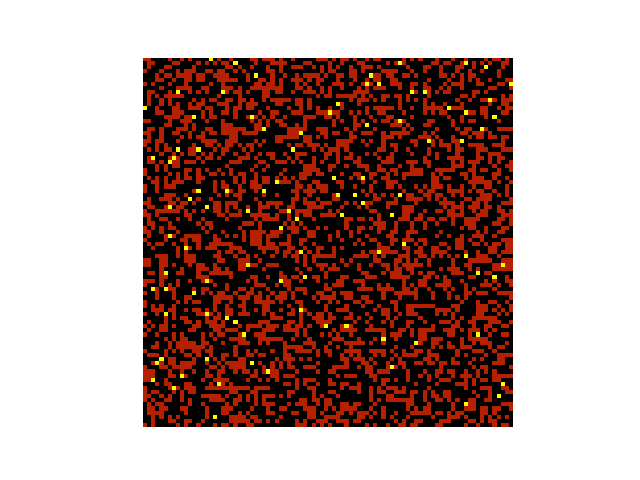}
\caption{t = 0}
\end{subfigure}
\begin{subfigure}[b]{.4\columnwidth}
\includegraphics[width=1.2\textwidth]{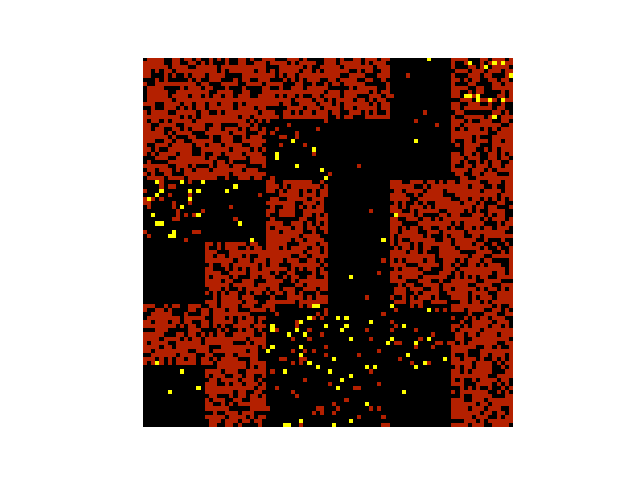}
\caption{t = 2}
\end{subfigure}

\begin{subfigure}[b]{.4\columnwidth}
\includegraphics[width=1.2\textwidth]{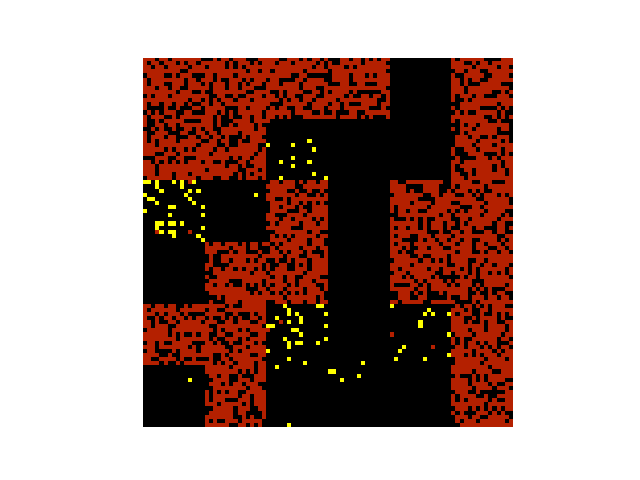}
\caption{t = 5}
\end{subfigure}
\begin{subfigure}[b]{.4\columnwidth}
\includegraphics[width=1.2\textwidth]{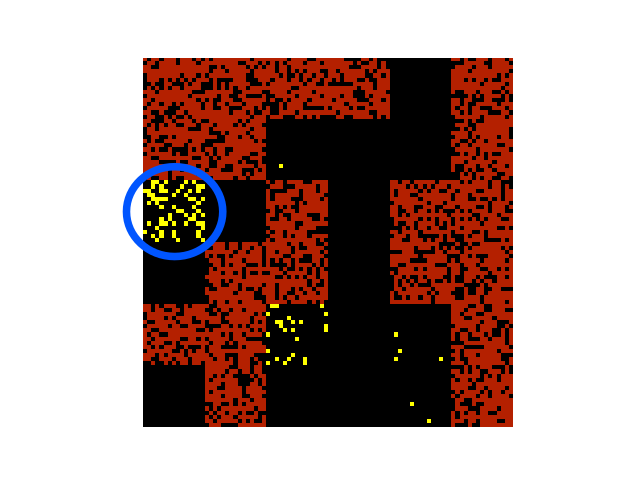}
\caption{t = 10}
\end{subfigure}

\begin{subfigure}[b]{.4\columnwidth}
\includegraphics[width=1.2\textwidth]{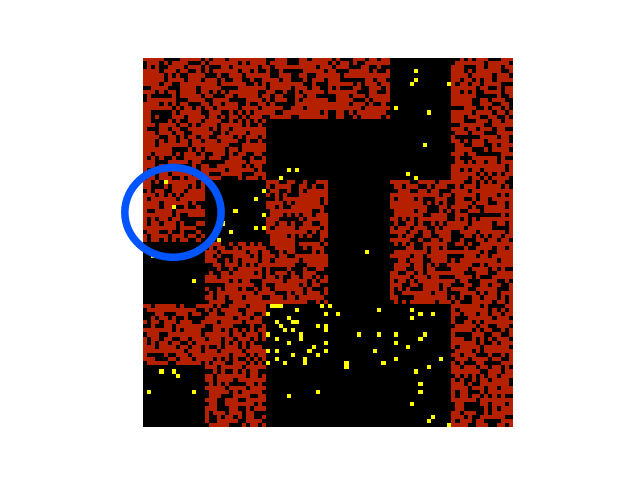}
\caption{t = 12}
\end{subfigure}
\begin{subfigure}[b]{.4\columnwidth}
\includegraphics[width=1.2\textwidth]{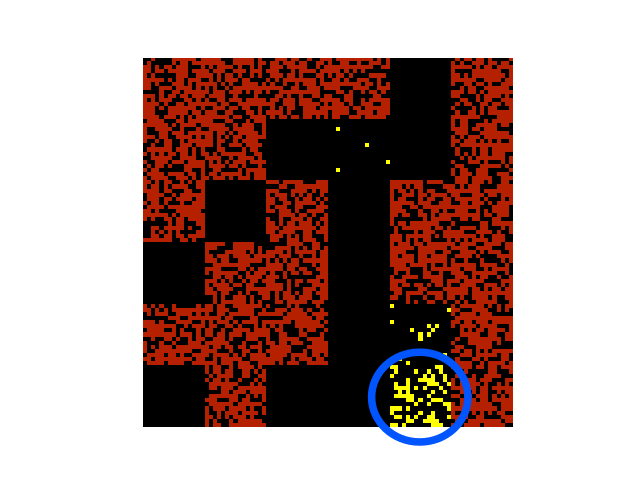}
\caption{t = 35}
\end{subfigure}

\begin{subfigure}[b]{.4\columnwidth}
\includegraphics[width=1.2\textwidth]{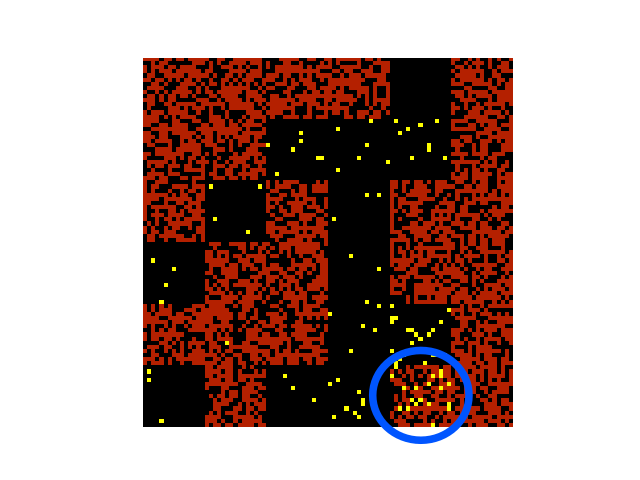}
\caption{t = 36}
\end{subfigure}
\begin{subfigure}[b]{.4\columnwidth}
\includegraphics[width=1.2\textwidth]{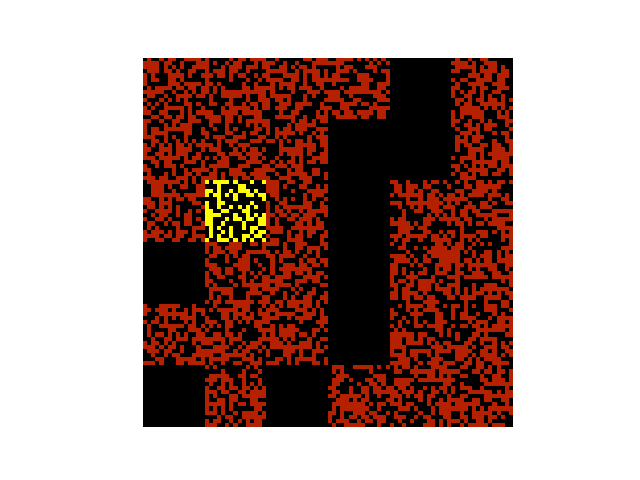}
\caption{t = 240}
\end{subfigure}
\begin{subfigure}[b]{0.7\columnwidth}
\includegraphics[width=\textwidth]{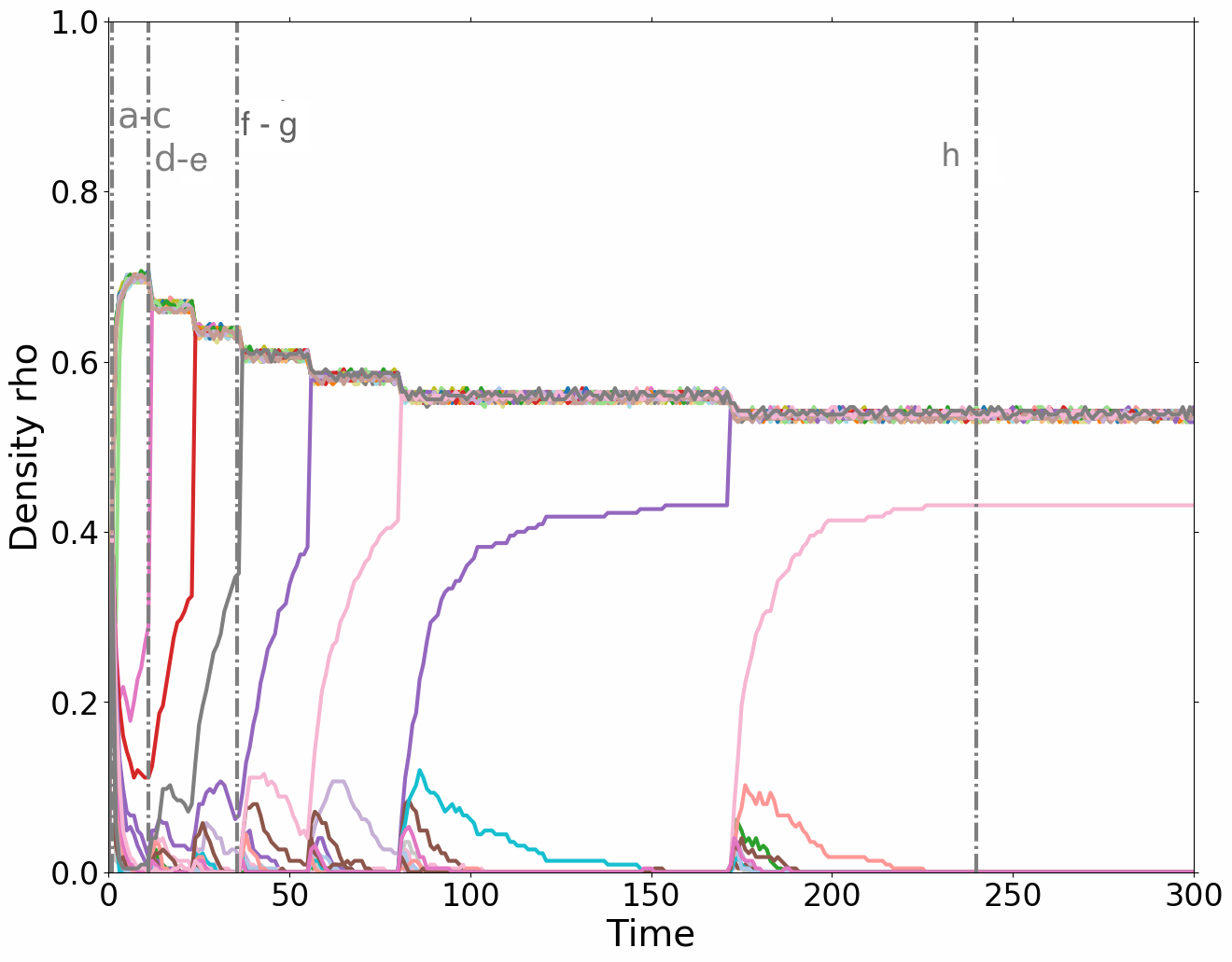}
\caption{Density profile}
\end{subfigure}
\caption{Evolution of the city for $p = 0.03, Q = 36$ and $H=225$. Panels (a-h) show the occupation of the different neighborhoods at different times. Egoists are represented in red, altruists in yellow, empty sites in black. (a) initial; (b) first steps; (c) usual segregation; (d-e): first invasion and altruist escape from the block surrounded in blue; (f-g): final invasion of the block surrounded in blue; (h): stationary state. In panel (i), each continuous line represents the evolution of the density of a single neighborhood. Vertical dashed lines show the times corresponding to panels (a-h).}
\label{fig:evolution}
\end{figure}

We now investigate how adding a fraction of altruists drives the system away from the frustrated pure egoist case to the optimal configuration observed in the pure altruist case. We find that, instead of a linear response, the system reaches the optimal state even at very low altruist concentrations ($p < 0.01$ in figure \ref{fig:palpha} a). To help understanding the origin of this strongly non-linear effect, the different panels of Fig.~\ref{fig:evolution} illustrate the evolution of a small system ($H = 225$, $Q = 36$ and $p = 0.04$). Initially, altruists (yellow) and egoists (red) are distributed randomly in the blocks (a), which all have a density $\rho \simeq \rho_0 = 0.4$. Then, blocks with the lowest densities are depleted by both altruists and egoists that prefer districts with higher densities. At some point, when the block density increases, the behavior of the two kinds of agents diverge. Altruists "sacrifice" themselves and leave these high density blocks, moving to blocks with lower densities, as this increases the utility of their many (former) neighbors, leading to an increase in global utility. On the other hand, egoists would loose individual utility by doing so, and therefore remain in these high density blocks which continue to feed on the remaining neighborhoods with $\rho < 1/2$. After a few iterations (Fig.~\ref{fig:evolution}b-c), selfish agents have gathered into "segregated" neighborhoods. This is the classical segregation observed in the pure egoist case \cite{pnas}, arising from the well studied amplification of density fluctuations. Note that all altruists have left the egoist blocks and gather into few blocks with lower densities (Fig.~\ref{fig:evolution}c) and then into a single neighborhood, whose density increases until it becomes attractive for egoist agents who "invade" it (Fig.~\ref{fig:evolution}d-e), while altruists leave it for other lower density blocks (Fig.~\ref{fig:evolution}e). The density of some of these new blocks then increases, allowing for successive egoist invasions (Fig.~\ref{fig:evolution}f-g). These migrations of egoist agents reduce the density of the overcrowded egoist blocks, increasing the overall utility. Eventually, the system reaches a stationary state in which no agent can move to increase its objective function (Fig.~\ref{fig:evolution}h).

\paragraph{Quantitative description.}
We now give a quantitative explanation of the decrease of egoist block densities and show that an altruist concentration $p \simeq 1/Q$ is sufficient to drive the system towards the optimal state, $\tilde{u} = 1$.
To understand altruists' dynamics, it is useful to replace their dynamics by an equivalent egoist dynamics with a utility $u_{\rm altr}(\rho)$ that differs from the original utility $u(\rho)$.
An exact mapping can be done in the following way. As mentioned above, each altruist agent tries to maximize the global utility $U = H \sum_{q=1}^Q \rho_q u(\rho_q)$.
In contrast, an egoist agent acts to maximize the link function $L=\sum_q \ell(\rho_q)$, with $\ell(\rho_q)$ given in Eq.~(\ref{eq:def:link}). As a result, an altruist agent exactly behaves as an equivalent egoist agent with a utility function $u_{\rm altr}(\rho)$ satisfying the relation
\begin{equation}
\rho u(\rho) = \int_0^{\rho} u_{\rm altr}(\rho') \, d\rho'
\end{equation}
since the resulting function to be maximized is the same. Differentiating this last equation, one finds
\begin{equation}
\label{eq:u_altr}
u_{\rm altr}(\rho)=\frac{\partial\bigl(\rho\, u(\rho)\bigr)}{\partial\rho}=
\begin{cases}
4\rho,  &\mbox{for }\rho\leq\frac{1}{2} 
\\ 2 (1 - 2\rho),  &\mbox{for }\rho>\frac{1}{2}
\end{cases}
\end{equation}

\begin{figure}%
\centering
\includegraphics[width=.65\columnwidth]{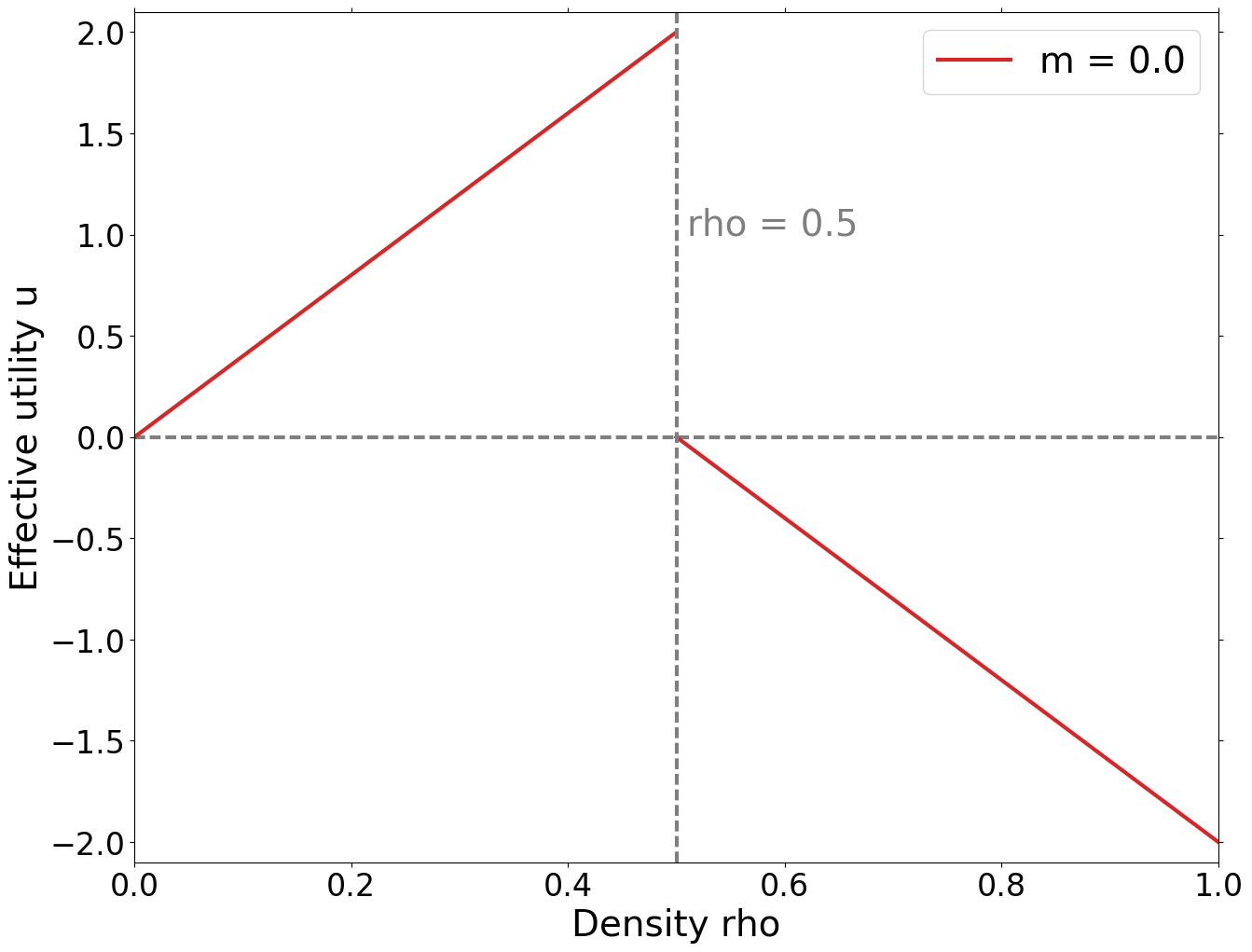}
\caption{\textit{Effective} utility function of altruistic agents.}
\label{fig:u_altr}
\end{figure}

\noindent
This effective utility function for altruists is plotted on Fig.~\ref{fig:u_altr}.
Note that this effective utility is {\it not} the one used to compute average or global utilities, it only helps understanding altruists' moves, since an altruist moves to a new block only if $u_{\rm altr}(\rho)$ increases. Fig.~\ref{fig:u_altr} shows that altruists have a clear preference for blocks with densities just below 1/2. The large discontinuity at $\rho = 1/2$ arises because at this density the original utility function $u(\rho)$ changes slope and starts to decrease. Then, an altruist moving from a block with $\rho < 1/2$ to a slightly more populated one with $\rho > 1/2$ induces a large decrease of total utility, since all its former neighbors loose utility (as the density of the initial block decreases) and so do its new neighbors, as the density of their block increases.

Fig.~\ref{fig:palpha}b suggests that the transition towards the optimal state is continuous and takes place at an altruist concentration $p \simeq 1/Q$ for all values of $Q$. This $Q$ dependence is important, since in the thermodynamic limit ($Q \rightarrow \infty$), the transition would take place at $p \rightarrow 0$. We now derive this result in a simple way by computing analytically the evolution of the average utility as a function of the altruist concentration $p$. Let's start with very low altruist concentrations and assume that the initial dynamics is dominated by egoists, which form the usual Schelling's overcrowded blocks, as observed above (Fig.~\ref{fig:evolution}c) and in previous work \cite{pnas}. Therefore, we take as starting point a city composed of $n_E$ egoist blocks with uniform density $\rho_e = \rho_E > 1/2$, such that $\rho_E =  (1 - p) Q \rho_0 / n_E$. Taking a uniform $\rho_e$ value is justified because any density fluctuation for $\rho_e > 1/2$ is rapidly wiped out by the dynamics, as shown by the unique density of egoist blocks in Fig.~\ref{fig:evolution}i. Altruists can be initially somewhat scattered over the remaining blocks but, as their effective utility clearly shows (Fig.~\ref{fig:u_altr}), they rapidly aggregate into a single block, leading to an altruist density $\rho_a = p Q \rho_0$ provided $\rho_a<1/2$, or equivalently
\begin{equation}
p < p_{\rm high} \equiv \frac{1}{2\rho_0 Q} \,.
\end{equation}
The driving force for the transition are the relative values of agents' utilities in egoist and altruist blocks, respectively $u_e = u_E = 2 - 2 \rho_E$ and $u_a = 2 \rho_a$ since $\rho_a < 1/2$ and $\rho_E > 1/2$. For very low $p$ values, $\rho_a$ is small, leading to $u_e > u_a$ and the system remains in the usual frustrated Schelling egoist state $\tilde{u}(p) \simeq \tilde{u}_E$ which is essentially constant. When $p$ reaches a value $p_{\rm low}$ such that $u(\rho_a + 1/H) > u(\rho_E) $, a first egoist can improve its utility by moving into the altruist block, whose density becomes $\rho_a + 1/H$ (Fig.~\ref{fig:evolution}d-f). This gives :
\begin{equation}
p_{\rm low} \equiv \frac{1-\rho_E - 1/H}{\rho_0 Q} \,,
\end{equation}
The density of the invaded block rapidly increases (Fig.~\ref{fig:evolution}e) and eventually reaches $1/2$. At this point, altruists' effective utility becomes negative, pushing them to leave for other lower density blocks (Fig.~\ref{fig:evolution}f). As previously, altruists gather in another single block of identical density $\rho_a = p Q \rho_0$. The invasion has led to a slight decrease of the density of egoist blocks to $\rho_e < \rho_E$, and therefore to a slight increase of egoists' utility, $u_e = u(\rho_e) > u(\rho_E)$. Successive invasions of the block partially filled by altruists are possible until $\rho_e$ decreases down to the value $\rho_e^*$ such that $u(\rho_e^*) = u(\rho_a + 1/H)$. This leads to $\rho_e^* = 1 - p Q \rho_0 - 1/H $ ($\rho_e^* > 1/2$ as long as $p<p_{\rm high}$). The equality of utilities implies $\tilde{u}(p)=u(\rho_a)=2pQ\rho_0 + 2/H$. 
When $p=p_{\rm high}$, the final (lowest) egoist density reaches the optimal value $\rho_e^* = 1/2$ and no further improvement in average utility is possible: $\tilde{u}(p) = 1$ (to simplify the discussion, we ignore here corrections of order $1/H$ that depend on the parity of H). This description remains valid for larger altruist concentrations, the only difference being that, at the end, the additional altruists form stable blocks with densities $\rho_a = 1/2$.

In summary, the evolution of the average utility $\tilde{u}$ follows:
\begin{equation}
\begin{cases}
\tilde{u}(p) = 2 - 2\rho_E &\mbox{for } p \leq p_{\rm low}
\\ \tilde{u}(p) = 2 p Q \rho_0 + 2/H &\mbox{for } p_{\rm low} \leq p \leq  p_{\rm high}
\\ \tilde{u}(p) = 1 &\mbox{for } p \geq p_{\rm high}
\end{cases}
\end{equation}
Our analysis predicts that plotting $\tilde{u}$ as a function of the rescaled altruist proportion
$p^* = p/p_{\rm high} = 2 p Q \rho_0$ should lead to a universal transition starting at $p^* = 2- 2\rho_E \simeq 0.586$ and ending at $p^* = 1$. Simulations perfectly confirm our calculations (Fig.~\ref{fig:palpha}b). 

\paragraph{Discussion.}
Our model illustrates the complexity of the dynamics produced by two types of agents, even when they follow simple rules. Introducing altruists into a population dominated by egoists increases the average utility much more rapidly than expected from a linear projection. The interplay between the different behaviors leads to complex "catalytic" phenomena. By catalytic, we mean that altruists are not "consumed" once they coordinate egoists, and can continue to help egoists finding the optimal configuration indefinitely. The global utility increase {\it per} altruist can be computed easily: $\delta U_{\rm altr} \equiv (U (p) - U (p=0)) / N_A  \simeq (1-0.56) \rho_0 Q H / (p \rho_0 Q H) =  0.44 / p$. When $p = 1/Q$, $\delta U_{\rm altr} \simeq 0.44 Q$. Each altruist induces a utility change proportional to the system size, which becomes infinite for infinite systems. 

Interestingly, while the stationary state of a system composed of a single type of agents (either egoists or altruists) can be mapped to an equilibrium state, this is no longer the case when including two types of agents, except if some restrictive conditions are met \cite{pnas}. In a thermodynamic analogy, the utility function can be mapped (in the zero temperature limit considered here) to a chemical potential, as shown in \cite{epl}, when a single type of agents is present. 
If a system with both egoist and altruist agents could be mapped to an equilibrium system, chemical potentials could be defined as $\mu_e(\rho_a,\rho_e)=u(\rho_a+\rho_e)$ and $\mu_a(\rho_a,\rho_e)=u_{\rm altr}(\rho_a+\rho_e)$. As chemical potentials derive from a free energy, their cross derivatives would be equal, $\partial \mu_e/\partial{\rho_a} = \partial \mu_a/\partial{\rho_e}$, leading to $u'(\rho)=u_{\rm altr}'(\rho)$.
This equality is not satisfied as seen from Figs.~\ref{fig:utility_function} and \ref{fig:u_altr}, showing that the system reaches a non equilibrium steady state.

We are well aware that simple models do not allow to draw any rigorous conclusion about what is going on in the {\it real} world \cite{jasss,ostrom,seuil}. While Schelling's segregation model neatly shows that one cannot logically deduce individual racism from global segregation, it may well be that for some towns racism is one cause of segregation, for some others not; at any rate the reasons behind urban segregation are far more complex than those that any simple model can come up with. Simple models can be helpful to analyze some interesting phenomena, the origin of which may be obscured in more complicated realistic settings. Ours may help thinking about the effectiveness of coordination by an infinitesimal proportion of altruist agents, but it cannot be directly applied to real systems. Real agents do not behave like these virtual robots: they are able to put their actions into context, to anticipate the behavior of the others and moreover, they disagree about what is the social "optimum" \cite{seuil,latour}.

\bibliography{refs}
\bibliographystyle{unsrt} 

\end{document}